\documentclass[preprint,showpacs,floatfix,amsmath,amssymb]{revtex4}
\usepackage{graphicx}
\usepackage{dcolumn}
\usepackage{bm}
\usepackage{amssymb}
\usepackage{graphicx}
\usepackage{amsmath}
\usepackage{xspace}
\begin{document}

\title{The effect of varying Fe-content on transport properties of K intercalated iron selenide
K$_{x}$Fe$_{2-y}$Se$_{2}$}
\author{D. M. Wang, J. B. He, T.-L. Xia, and G. F. Chen}
\email{genfuchen@ruc.edu.cn}

\affiliation{Department of Physics, Renmin University of China,
Beijing 100872, P. R. China}

\date{\today}

\begin{abstract}
We report the successful growth of high-quality single crystals of
potassium intercalated iron selenide K$_{x}$Fe$_{2-y}$Se$_{2}$ by
Bridgeman method. The effect of iron vacancies on transport
properties was investigated by electrical resistivity and magnetic
susceptibility measurements. With varying iron content, the system
passes from semiconducting/insulating to superconducting state.
Comparing with superconductivity, the anomalous ``hump" effect in
the normal state resistivity is much more sensitive to the iron
deficiency. The electrical resistivity exhibits a perfect metallic
behavior (R$_{300K}$/R$_{35K}$$\approx$42) for the sample with
little iron vacancies. Our results suggest that the anomalous
``hump" effect in the normal state resistivity may be due to the
ordering process of the cation vacancies in this
non-stoichiometric compound rather than magnetic/structure
transition. A trace of superconductivity extending up to near 44 K
was also detected in some crystals of K$_{x}$Fe$_{2-y}$Se$_{2}$,
which has the highest T$_c$ of the reported iron selenides.

\end{abstract}


\maketitle

The discovery of superconductivity in iron-based pnictides has
attracted a great deal of research interest.\cite{Ishida} The
PbO-type $\alpha$-FeSe$_{x}$, which has an extremely simple
structure with only FeSe$_{4}$ tetrahedral layers stacked along
c-axis, the same as the FeAs$_{4}$ tetrahedral layers found in
pnictides, was discovered subsequently at 8 K in samples prepared
with Se deficiency.\cite{MKWu} The superconductivity of FeSe$_{x}$
can be further enhanced up to 14 K by partially replacing Se with
Te.\cite{Sales} High pressure studies at 4.5 GPa have dramatically
yielded the superconductivity at 37 K.\cite{Medvedev} Very
recently, by intercalating potassium between the FeSe layers,
i.e., K$_{x}$Fe$_{2}$Se$_{2}$, superconductivity has been observed
up to 30 K,\cite{XLChen} which is the same as those in optimal
doped pnictides AFe$_{2}$As$_{2}$ (A=Ca, Sr, Ba, Eu).\cite{Ishida}
In iron pnictides, many open questions are waiting to be solved,
especially the relationship between superconductivity, structural
change and magnetic order. Hence, systematic investigations on the
superconductivity in K$_{x}$Fe$_{2}$Se$_{2}$ can provide us some
hints on the difference or common features between pnictide and
selenide systems, and further shed light on the mechanism of
superconductivity in Fe-based superconductors.

In present work, we report the successful growth of high quality
single crystals of K$_{x}$Fe$_{2-y}$Se$_{2}$ with different
Fe-deficiency using a Bridgeman method and a systematic study on
effect of Fe-deficiency on the transport properties. We find that
the system displays unique properties such as electrical
resistivity ranging from superconducting to
semiconducting/insulating, which can be tuned by adjusting the
growth conditions. Comparing with superconductivity, the anomalous
``hump" effect in the normal state resistivity in superconducting
samples is much more sensitive to the iron deficiency. We
speculate that the anomalous ``hump" effect in the normal state
resistivity may be due to the ordering process of the cation
vacancies in this non-stoichiometric compound rather than
magnetic/structure transition. Furthermore, a trace of
superconductivity extending up to near 44 K was detected in some
batches of K$_{x}$Fe$_{2-y}$Se$_{2}$, which has the highest T$_c$
of the reported iron selenides. The appropriate nominal
compositions and growth conditions are important key parameters to
modulate the transport properties in K$_{x}$Fe$_{2-y}$Se$_{2}$.

\begin{figure}[h]
\includegraphics[width=12 cm]{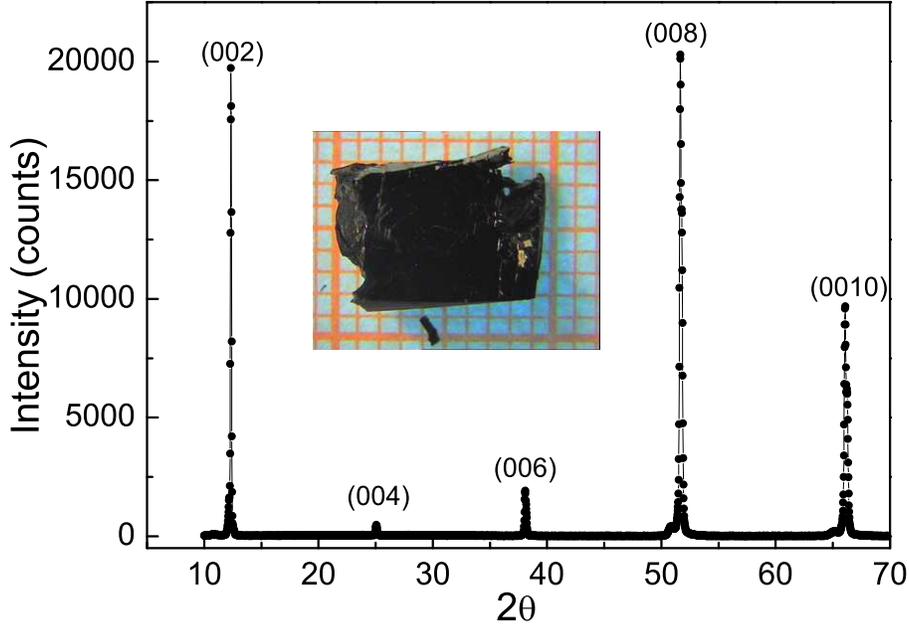}
\caption{(Color online) The X-ray diffraction pattern of
K$_{0.8}$Fe$_{2}$Se$_{2}$ crystal with mainly the (00$\ell$)
reflections indicates that the crystals are cleaved along the a-b
plane. Inset shows the typical photography of the grown
K$_{0.8}$Fe$_{2}$Se$_{2}$ crystal.}
\end{figure}

 All the samples were prepared by Birdgeman method. Fe$_{1+y}$Se
was firstly synthesized as precursor by reacting Fe powder with Se
powder at 750 $^{0}$C for 20 hours. K pieces and Fe$_{1+y}$Se
powder were put into an alumina crucible with nominal compositions
(Hereafter, all the compositions used in this paper refer to the
nominal compositions) as K$_{x}$Fe$_{2+y}$Se$_{2}$
($0.6$$\leq$x$\leq1.0$; $0$$\leq$y$\leq0.3$). The alumina crucible
was then sealed into Tantalum tube with Argon gas under the
pressure of 1.5 atom, then the sealed Ta tube were vacuum sealed
into quartz tubes. In some cases, the alumina crucible was
directly sealed into thick walled quartz tube with Ar under the
pressure of 0.4 atom. The sample was put in box/tube furnace and
heated to 1050 $^{0}$C slowly and held there for 2 hours, and then
was cooled to 750 $^{0}$C over 60$\sim$200 hours to grow the
single crystals. The obtained crystals with sizes up to
8mm$\times$8mm$\times$5mm have the form of platelets with shinny
surface. These crystals were characterized by X-ray diffraction
(XRD). The elemental composition of the single crystal was checked
by energy dispersive X-ray (EDX) spectroscopy analysis. The
resistivity was measured by a standard 4-probe method. The DC
magnetic susceptibility was measured with a magnetic field of 10
Oe. These measurements were performed down to 2 K in a Physical
Property Measurement System (PPMS) of Quantum Design with the VSM
option provided.

\begin{figure}[t]
\includegraphics[width=8 cm]{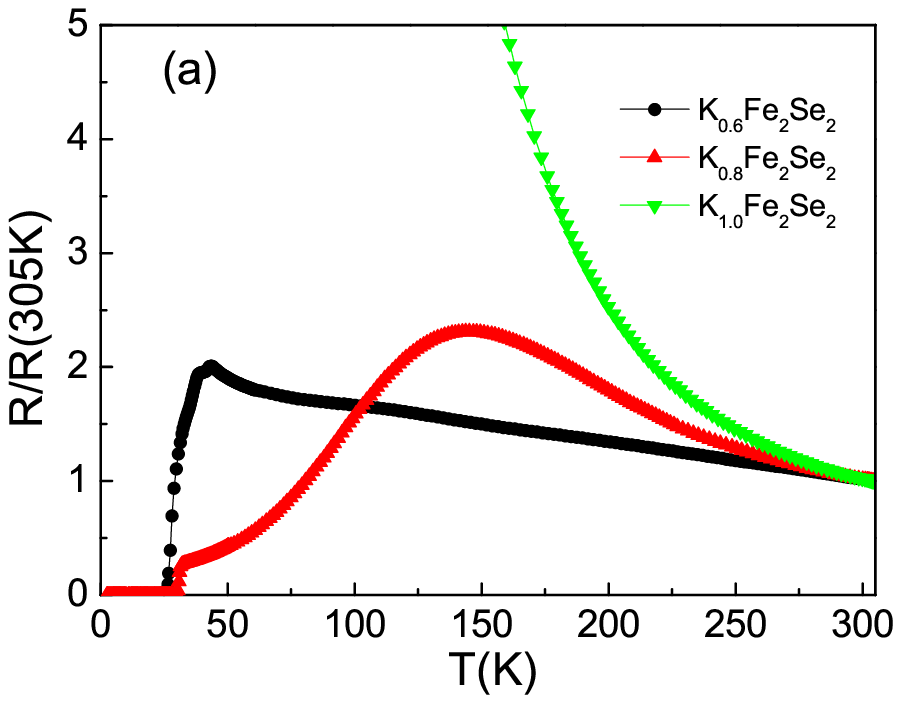}
\includegraphics[width=8 cm]{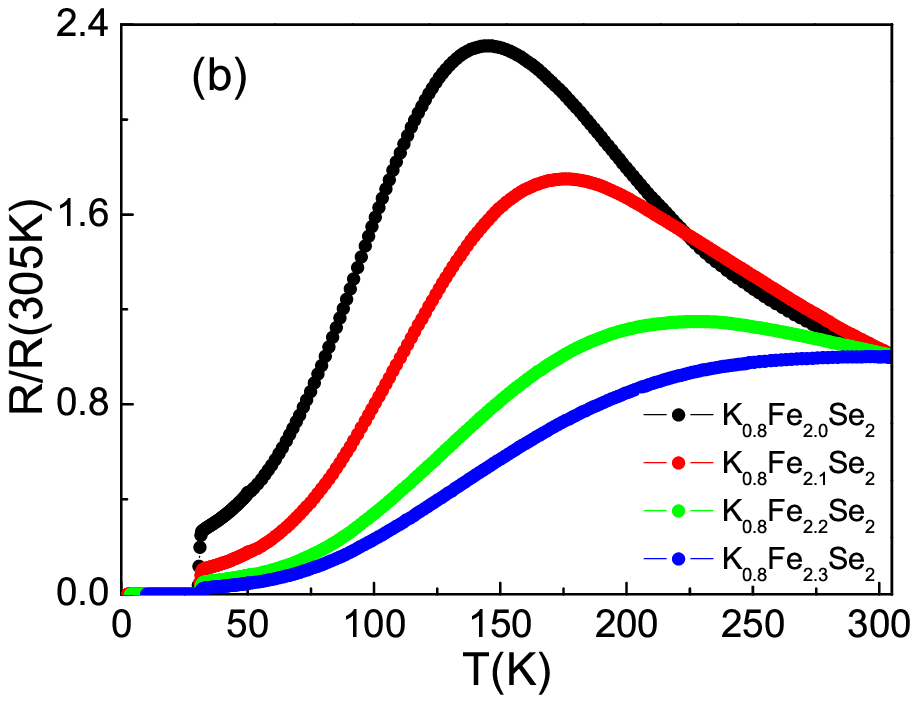}
\caption{(Color online) Temperature dependent resistivity of grown
K$_{x}$Fe$_{2+y}$Se$_{2}$ (0.6$\leq$x$\leq$1.0, 0$\leq$y$\leq$0.3)
crystals. Different electronic transport properties demonstrated
by the resistivity curves have been observed.}
\end{figure}

Figure 1 shows the X-ray diffraction pattern of
K$_{0.8}$Fe$_{2}$Se$_{2}$ (nominal composition) with the 00$\ell$
($\ell=2n$) reflections, which is consistent with previous
reports. The lattice constant c = 14.15 $\AA$ was calculated from
the higher order peaks, comparable to the result from powder
diffraction.\cite{XLChen} The picture of grown crystal with
nominal composition K$_{0.8}$Fe$_{2}$Se$_{2}$ is presented in the
inset of Fig.1.

Firstly, we investigated the relation between K concentration and
the behavior of the resistivity. Figure 2a shows clearly that with
the increase of nominal K doping, the system evolves from
semiconductor-like (superconducting) to a typical structure same
as previous report with a broad peak around 140 K, then becomes a
semiconductor/insulator-like system without superconducting
observed inside. The EDX data shows that as the nominal content of
K was increased from 0.8 to 1.0, the iron content in the obtained
single crystals systemically decreased from 1.8 to 1.5 and the
superconductivity finally disappeared. This indicates that the
high deficiency of Fe has a major detrimental effect on the
transport properties.

\begin{figure*}[t]
\includegraphics[width=16 cm]{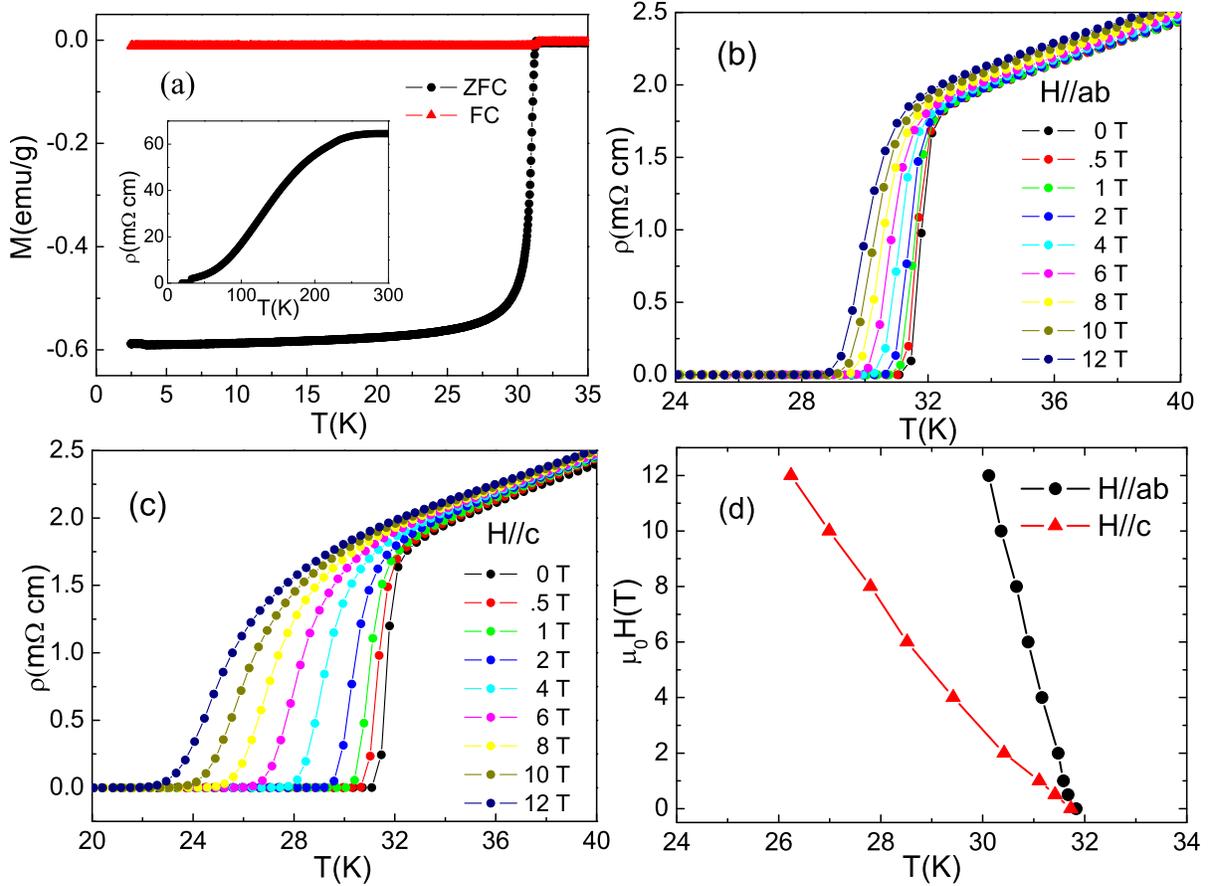}
\caption{(Color online) (a) DC magnetic susceptibility of
K$_{0.8}$Fe$_{2.3}$Se$_{2}$ crystal. Inset shows the temperature
dependence of resistivity between 300 K and 20 K; The temperature
dependence of resistivity for K$_{0.8}$Fe$_{2.3}$Se$_{2}$ crystal
with the applied field parallel (b) and perpendicular (c) to the ab
plane; (d) Temperature dependence of H$_{c2}$(T) for
K$_{0.8}$Fe$_{2.3}$Se$_{2}$ crystal.}
\end{figure*}

Secondly, we systematically changed the concentrations of Fe ions
with K fixed to study the transport behavior. Figure 2b shows the
temperature dependence of resistivity for
K$_{x}$Fe$_{2+y}$Se$_{2}$ crystals between 2 K and 300 K. It is
obvious that the behavior of the electrical transport has changed
with the increase of Fe concentrations (corresponding to the
decrease of Fe deficiency.) When the Fe deficiency is large, the
resistivity curve possesses an obvious metal-insulator (MI)
transition at around 140 K. At high temperatures, the sample shows
the semiconductor-like behaviors, while below the transition
temperature it presents a metallic behavior.\cite{XLChen} With
increasing the Fe concentration, the hump diminishes gradually and
the position shifts to high temperature, and finally vanishes in
the sample with nominal composition K$_{0.8}$Fe$_{2.3}$Se$_{2}$.
Here the electrical resistivity exhibits a perfect metallic
behavior. However, the superconducting critical temperature,
T$_{c}$, increases slightly (less than 0.8 K) with decreasing Fe
deficiency. We note that there is not any anomaly in the
temperature dependence of magnetic susceptibility and there is
also no structural phase transition occurring over the temperature
ranging from 60 to 300 K.\cite{LLSun} The semiconductor-metal like
transition may be attributed to the ordering process of the cation
vacancies in the non-stoichiometric compound of
K$_{x}$Fe$_{2-y}$Se$_{2}$ and it significantly influences the
electrical resistivity. We notice that similar issues concerning
order-disorder phenomena have been discussed in the transition
metal carbides and the high-T$_{c}$ compounds where the ordering
of the vacancies play an important role in controlling the
normal-state resistivity and superconductivity.\cite{LCDy,BWVeal}
Interestingly, the temperature dependence of electrical
resistivity under high pressure has been investigated by Sun et
al.,\cite{LLSun} where the ``hump" was successively suppressed by
applying pressure and the superconductivity was destroyed
simultaneously, in contrast to the results of our study. There is
no obvious correlation between the superconducting critical
temperature and the broad hump.

\begin{figure}[b]
\includegraphics[width=8 cm]{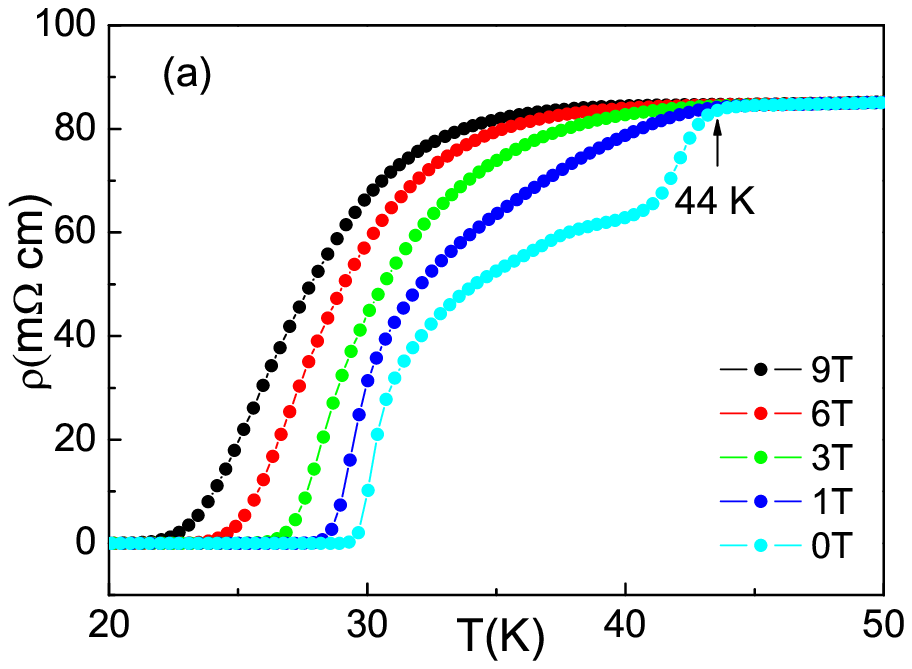}
\includegraphics[width=8 cm]{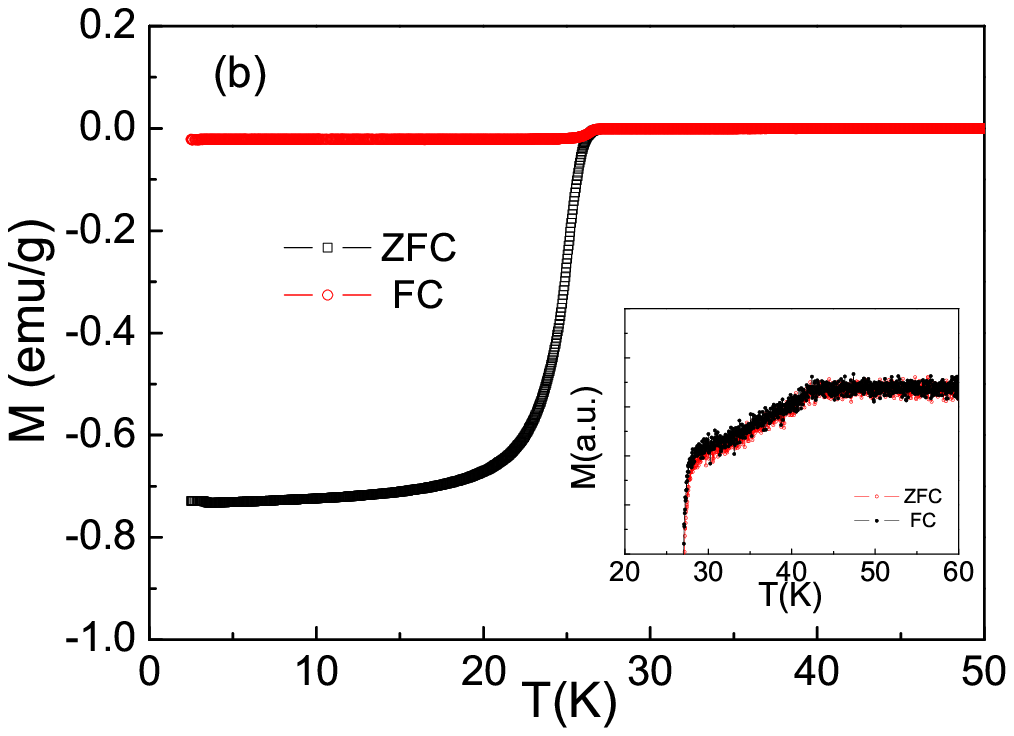}
\caption{(Color online) (a) Temperature dependence of the
resistivity under different magnetic field; (b) Magnetic
susceptibility of the the same samples. Inset of figure 4b is the
enlarged part from figure 4b which clearly indicates the transition
around 44 K.}
\end{figure}

Figure 3a presents the magnetic susceptibility as a function of
the temperature from 3 K to 35 K for the sample with nominal
composition K$_{0.8}$Fe$_{2.3}$Se$_{2}$ under the magnetic field
of 10 Oe. The zero field cooling (ZFC) and field cooling (FC)
susceptibility both show sharp transitions at around 32 K. From
ZFC, it can be estimated that the superconducting volume fraction
at 10 K is about 100$\%$, which demonstrates the bulk
superconductivity in the samples. Inset of Fig.3a shows the
typical resistivity curve of K$_{0.8}$Fe$_{2.3}$Se$_{2}$, which
clearly shows the metallic behavior with high residual resistivity
ratio (R$_{300K}$/R$_{35K}$$\approx$42).

Figure 3b and 3c show the behavior of resistivity of
K$_{0.8}$Fe$_{2.3}$Se$_{2}$ in external magnetic field up to 12 T.
In Figure 3b, the applied field is within the ab plane ($H\| ab$)
while in Fig.3c the applied field is parallel to the c axis ($H\|
c$). We see the superconducting transition is suppressed in both
conditions but the effect of magnetic field is much larger when
the field is applied along the c axis of the single crystals in
stead of within the ab plane. The corresponding upper critical
field $H_{c2}$ as a function of temperature obtained from a
determination of the midpoint of the resistive transition is
plotted in Fig.3d. The curves are steep with slopes
$-dH_{c2}/dT|_{T_{c}}$=7.20 T/K for $H\|ab$ and
$-dH_{c2}/dT|_{T_{c}}$=2.17 T/K for $H\|c$. According to the
Werthamer-Helfand-Hohenberg (WHH) formula\cite{WHH}
$H_{c2}(0)=-0.69(dH_{c2}/dT)T_{c}$ and taking 32 K as T$_{c}$, the
upper critical fields are estimated to be $H_{c2}^{ab}$=159 T and
$H_{c2}^{c}$=48 T. The ratio of $H_{c2}^{ab}/H_{c2}^{c}$ is about
3.3 which is consistent with previous reports.\cite{MHFang}

One issue we want to point out is that a trace of
superconductivity with T$_{c}$ up to 44 K has been observed in
several batches of samples with nominal composition
K$_{0.8}$Fe$_{2}$Se$_{2}$. Figure 4a shows the temperature
dependence of the electrical resistivity in the low temperature
region. It is clear that there exist two transition steps on the
resistivity curve measured under zero field, one at about 38 K and
another at about 44 K. To verify whether the transition at 44 K is
due to superconductivity, we measured the field dependence of
resistivity and observed the resistivity curves change at around
44 K, and the kink shifts to low temperature with increasing
magnetic fields. Magnetic susceptibility has also been measured on
the same samples and it is clearly shown in the Fig.4b that once
the temperature is lower than around 44 K, the magnetization
begins to decrease and the rate of decreasing greatly enhances
once the temperature is lower than 28 K, which corresponds to the
transition temperature of zero resistivity. Hence it raises one
question: where does the superconductivity come from, ``122" phase
with super high quality, or other new phase? Detailed
characterization of the superconducting phase is in progress.

In conclusion, we have successfully grown series of single
crystals K$_{x}$Fe$_{2-y}$Se$_{2}$ with different Fe vacancies. A
trace of superconductivity extending up to near 44 K was also
observed in some K$_{x}$Fe$_{2-y}$Se$_{2}$ crystals, which has the
highest T$_c$ of the reported iron selenides. The anomalous
semiconductor-metal-like transition is observed only in the sample
with high level of Fe vacancies. We speculate that the anomaly is
associated with the ordering of the framework vacancies, which
significantly influences the electrical resistivity. The more
ordered phase showing a large reduction in residual resistivity is
due to reduction of electron-vacancy scattering. A more detailed
study is needed in order to better understand the correlation of
superconductivity and the anomalous metal-insulator transition.


\begin{thebibliography}{24}

\bibitem{Ishida} K. Ishida, Y. Nakai, and H. Hosono, J. Phys. Soc. Jpn.
\textbf{78}, 062001 (2009).

\bibitem{MKWu} F. C. Hsu,
J. Y. Luo, K. W. The, T. K. Chen, T. W. Huang, P. M. Wu, Y. C. Lee,
Y. L. Huang, Y. Y. Chu, D. C. Yan and M. K. Wu, Proc. Nat. Acad.
Sci. {\bf 105}, 14262 (2008).

\bibitem{Sales} B. C. Sales, A. S. Sefat, M. A. McGuire, R. Y. Jin, D. Mandrus, and
Y. Mozharivskyj, Phys. Rev. B {\bf 79}, 094521 (2009).

\bibitem{Medvedev} S. Medvedev, T. M. McQueen, I. A. Troyan, T. Palasyuk, M. I.
Eremets, R. J. Cava, S. Naghavi, F. Casper, V. Ksenofontov, G.
Wortmann, and C. Felser, Nat. Mater. {\bf 8}, 630 (2009).

\bibitem{XLChen}
J. G. Guo, S. Jin, G. Wang, S. Wang, K. Zhu, T. Zhou, M. He and X.
L. Chen, Phys. Rev. B {\bf 82}, 180520 (2010).

\bibitem{LLSun} J. Guo, L. L. Sun, C. Zhang, J. G. Guo, X. L. Chen,
Q. Wu, D. C. Gu, P. W. Gao, X. Dai, and Z.-X. Zhao, arXiv:1101.0092.

\bibitem{LCDy} L. C. Dy, Wendell. S. Williams, J. Appl. Phys. \textbf{53},
8915 (1982).

\bibitem{BWVeal} B. W. Veal, H. You, A. P.Paulikas, H. Shi, Y. Fang,
and J. W. Downey, Phys. Rev. B 42, 4770 (1990).

\bibitem{WHH} N. R. Werthamer, E. Helfand and P. C. Hohenberg, Phys. Rev. \textbf{147}, 295(1966).

\bibitem{MHFang} M. H. Fang, H. D. Wang, C. H. Dong,
Z. J. Li, C. M. Feng, J. Chen, H. Q. Yuan. arXiv:1012.5236.


\end{thebibliography}
\end{document}